\documentstyle[12pt,psfig]{article}
\def\Oc{\Omega_{ccc}}
\def\als{\mbox{$\alpha_s$}}
\def\MA{{\cal M}}
\def\M0{{\cal M}_o}

\begin{document}

\baselineskip=19pt

\begin{center}

{\bf $\Omega_{ccc}$ production via fragmentation at LHC}

Saleev V.A.

Samara State University, Samara, Russia

\end{center}

\begin{abstract}
In the framework of the leading order of perturbative QCD
and the nonrelativistic quark-diquark model of baryons
we have obtained fragmentation function for c-quark to split
into $\Oc$ baryon.
It is shown that at LHC one can
expect $3.5\cdot 10^3$ events with $\Omega_{ccc}$  at
$p_{\bot}>5$ GeV/c and $\vert y\vert <1$ per year.
\end{abstract}

\subsection*{1. Introduction}
During last years the physics of baryons containing heavy quarks
was a subject of increasing attention \cite{1}. The new interesting
data on heavy baryon production rates and decay widths have been
obtained \cite{2}. There are theoretical predictions for doubly heavy
baryon production cross sections \cite{3}-\cite{6}. It is expected,
that $\Xi_{cc}$ and $\Xi_{cc}^*$ baryons, containing two charmed
quarks, can be detected already at the Tevatron collider ($\sqrt
{s}=1.8$ TeV), and at LHC ($\sqrt{s}=14$ TeV) their production rate
increase by $10^4$ times \cite{7}. The predicted numbers of (bc)- and
(bb)-baryons for LHC are 1/3 and 1/100, correspondingly, of the
number of (cc)-baryons \cite{8}.

The purpose of this paper is to estimate the production rate for $\Oc$
baryons at LHC via fragmentation. Most of the fragmentation functions
for heavy quarkonia and doubly heavy diquarks have now been calculated
\cite{3,9,11}. Using the same approach we calculate here fragmentation
function for c-quark to split into $\Oc$ baryon, which is considered
as a nonrelativistic system of c-quark and (cc)-diquark \cite{6,10,10b}.

\subsection*{2. Charmed quark fragmentation into $\Oc$}
The fragmentation function is given by the formula \cite{11}:
\begin{equation}
D_{c\to\Oc}(z,\mu_0)=\frac{1}{16\pi^2}\int_{s_{min}}^{\infty}ds
\lim_{q_0\rightarrow\infty}
\frac{\vert\MA\vert^2}{\vert\M0\vert^2},
\end{equation}
where $\MA$ is the amplitude for the producing $\Oc$ plus
$(\bar c\bar c)$-diquark
with the total 4-momentum $q=(q_0,0,0,q_3)$
and invariant mass  $s=q^2$,
$\M0$ is the amplitude to create on-shell c-quark with the same
3-momentum $\vec q$ from the same source.
The lower limit in the integral is:
$$ s_{min}=\frac{M^2}{z}+\frac{m_{cc}^2}{1-z},$$
where
$$z=\frac{p_0+p_3}{q_0+q_3},$$
$p=(p_0,0,0,p_3)$ is the $\Oc$ baryon 4-momentum. In the axial gauge
for the gluon propagator associated with four-vector $n=(1,0,0,-1)$
$$
d_{\mu\nu}(k)=-g_{\mu\nu}+\frac{k_{\mu}n_{\nu}+k_{\nu}n_{\mu}}{(kn)},$$
the fragmentation contribution comes only from the Feynman diagram
shown in Fig.1:
\begin{eqnarray}
\MA&=&\frac{\vert\Psi_{\Oc}(0)\vert}{\sqrt{2m_{cc}}}
\frac{4\delta^{ij}}{3\sqrt{3}}(4\pi\als)^2
\frac{F_D(k^2)}{k^2(s-m_c^2)}\bar\Psi^{\beta}(p)\gamma_{\nu}(\hat
q+m_c)\hat G
\nonumber\\
&&
d^{\mu\nu}(k) V_{\alpha\mu\beta}(q,p_c)\varepsilon^*_{\alpha}(q'),
\end{eqnarray}
where $\hat G$ describes the creation of c-quark with the total
4-momentum $q=p+q'$, $4\delta^{ij}/3\sqrt{3}$ is the color factor of
the diagram, $F_D(k^2)$ is the vector diquark form factor for the
vertex $g^*\to (cc)+(\bar c\bar c)$,$k=q'+p_{cc}$, $\bar\Psi^{\beta}(p)$

is the spin-vector of the spin-3/2 baryon,
$\varepsilon^*_{\alpha}(q')$ is the polarization four-vector of
$(\bar c\bar c)$-diquark, $\Psi_{\Oc}(0)$ is the nonrelativistic $\Oc$
wave function at the origin in the quark-diquark approximation.
The vertex $g^*\to (cc)+(\bar c\bar c)$ is written as follows
\begin{equation}
-ig_sT^aV_{\alpha\mu\beta}(q',p_{cc})F_D(k^2),
\end{equation}
where
$$
V_{\alpha\mu\beta}(q',p_{cc})=
-g_{\alpha\beta}(q'-p_{cc})_{\mu}-g_{\beta\mu}(3p_{cc}+2q')_{\alpha}
+g_{\mu\alpha}(3q'+2p_{cc})_{\beta},
$$
\begin{equation}
F_D(k^2)=F_{D0}\left (\frac{m_{cc}^2}{k^2}\right )^2,
\end{equation}
$$F_{D0}=128\pi\als\frac{\vert \Psi_{cc}(0)\vert ^2}{m_{cc}^3},$$
$\Psi_{cc}(0)$ is the nonrelativistic (cc)-diquark wave function
at the origin, $m_{cc}$ is the diquark mass.

In the nonrelativistic approximation one has $p_{cc}=(1-r)p$ and
$p_c=rp$, where $r=m_c/M$.
Scalar products of $k$, $p$ and $q$ are presented as follows:
$$k^2=(1-r)(s-m_c^2),\quad 2(qk)=(2-r)(s-m_c^2),$$
$$2(kp)=s-m_c^2,\quad 2(pq)=s-m_c^2+2rM^2. $$
After summation over polarizations of the squared amplitudes and
integration over s we have found the fragmentation function at the
scale $\mu_0=4m_c$:
\begin{equation}
D_{c\to\Oc}(z,\mu_0)=\frac{\vert\Psi_{\Oc}(0)\vert^2}{M^3}
\als^2(\mu_0)F^2_{D0}\Phi_c (z),
\end{equation}
\begin{eqnarray}
\Phi_c(z)&=&\frac{36z^4(1-z)^3}{35(z-3)^{14}}
(113519 z^8-1303182 z^7+8764206 z^6-26818758 z^5\nonumber\\
&&+52452396 z^4
-73464138 z^3+66215394 z^2-32322402 z+6506325).\nonumber
\end{eqnarray}

The recalculating of the fragmentation function from the starting
point $\mu_0$ to $\mu>\mu_0$ may be done using evolution equation
\begin{equation}
\mu\frac{\partial D}{\partial \mu}(z,\mu)=\int_z^1\frac{dy}{y}
{\cal P}_{c\to c}\left (\frac{z}{y},\mu\right ) D(y,\mu),
\end{equation}
where ${\cal P}_{c\to c}(x,\mu)$ is the splitting function at the
leading order in $\als$
\begin{equation}
{\cal P}_{c\to c}(x,\mu)=\frac{4\als (\mu)}{3\pi}\left
(\frac{1+x^2}{1-x}\right )_+,
\end{equation}
$$ f(x)_+=f(x)-\delta (1-x)\int_0^1 f(x')dx'.$$
Fig. 2 shows fragmentation function $D_{c\to\Oc}(z,\mu)$
normalized to unity at the $\mu=\mu_0$ (curver 1)
and $\mu=45$ GeV (curver 2).
We have obtained that average fractions of the $\Oc$ momentum
are equal to $<z>_{\mu_0}=0.61$ and $<z>_{45}=0.42$. Since, one has at
leading order in $\als$
$$ \int_0^1{\cal P}_{c\to c}(x,\mu)dx=0,$$
the evolution equation implies that the fragmentation probability
$P_{c\to\Oc}$ don't evolve with the scale $\mu$:
\begin{equation}
P_{c\to\Oc}= \int_0^1  D_{c\to\Oc}(z,\mu_0)dz=
A_c\als^2(\mu_0) F_{D0}^2\frac{\vert\Psi_{\Oc}(0)\vert^2}{M^3},
\end{equation}
where
$$A_c=\frac{263585448}{5}\ln\left (\frac{3}{2}\right )-
\frac{1100381933317}{51480}\approx 4.19\cdot10^{-3}.$$

We used for numerical calculation computer program "schroe" \cite{12}
with the power-low quark-quark
interaction potential from \cite{13}:
$$V_{Q\bar Q}(r)=-A+B(r\cdot 1\mbox{ GeV})^n,\quad
V_{QQ}=\frac{1}{2}V_{Q\bar Q},$$
where $A=-8.064$ GeV, $B=6.898$ GeV and
$n=0.1$.
We have obtained masses of the vector (cc)-diquark,
$\Oc$ baryon and values of nonrelativistic wave functions of
(cc)-diquark and $\Oc$ baryon at the origin:
$m_{cc}=3.48$ GeV, $M=4.70$ GeV, $\vert\Psi_{cc}(0)\vert^2=0.03$
GeV$^3$,
$\vert\Psi_{\Oc}(0)\vert^2=0.115$ GeV$^3$.
Using above set of parameters with $\als$=0.3,
we have found the fragmentation probability
\begin{equation}
       P_{c\to \Oc}=1.15\cdot 10^{-9}.
\end{equation}

\subsection*{3. Hadronic production of $\Oc$ baryons.}

Accordingly to factorization approach, $p_{\bot}$-spectrum of $\Oc$
baryons in hadronic collisions at high $p_{\bot}$ may be presented
as the convolution of c-quark
$p_{\bot}$-spectrum and fragmentation function $D_{\Oc}(z,\mu)$
at  the scale $\mu\approx \sqrt{p^2_{\bot}+M^2}$:
\begin{equation}
\frac{d\sigma}{dp_{\bot}}(p\bar p\to\Oc X)=\int_0^1D_{c\to\Oc}(z,\mu)
\frac{d\sigma}{dp_{\bot c}}\left (p\bar p\to c X, p_{\bot
c}=\frac{p_{\bot}}{z}\right )dz,
\end{equation}
\begin{eqnarray}
\frac{d\sigma}{dp_{\bot}}(p\bar p\to c
X)&=&2p_{\bot}K\int_{y_{min}}^{y_{max}}dy\int_{x_{1,min}}^1
dx_1\frac{x_1x_2s}{x_1s+u-m_c^2}
\nonumber\\
&&\sum_{i,j}F^p_i(x_1,\mu)F^{\bar p}_j(x_2,\mu)
\frac{d\hat\sigma}{d\hat t}(ij\to c\bar c),
\end{eqnarray}
where $F_i^{p,\bar p}(x_{1,2},\mu)$ are quark or gluon distribution
functions in proton (antiproton),
$d\hat\sigma/d\hat t (ij\to c\bar c)$ are differential cross sections
for partonic subprocesses $gg\to c\bar c$, $q\bar q\to c\bar c$,
$K\approx 3$ is the phenomenological
factor, which takes into account the next to leading order contribution
in $\als$. In the calculation we have used CTEQ5 \cite{14}
parameterization for distribution functions and $K$-factor have been
fixed using data for b-quark production in $p\bar p$-collisions at
$\sqrt{s}=1.8$ TeV.

After numerical integration of (10)-(11) at $p_{\bot}>5$ GeV/Â and
$\vert y\vert <1$ we have found:
\begin{equation}
\sigma (p\bar p\to \Oc, \sqrt{s}=1.8 \mbox{ TeV})=3.6\cdot 10^{-3}
\mbox{ pb},
\end{equation}
\begin{equation}
\sigma (p\bar p\to \Oc, \sqrt{s}=14 \mbox{ TeV})=3.45\cdot 10^{-2}
\mbox{ pb}.
\end{equation}
For the Tevatron, with an integrated luminosity of
$\sim 10^2$ pb$^{-1}/yr$ , we predict smaller than one event per year,
while the LHC, with luminosity $\sim 10^5$ pb$^{-1}/yr$,
should produce of order $3.5\cdot 10^3$ events per year.

In conclusion, the calculation of the production of the hadrons
containing heavy quarks
in the fragmentation approach gives rather lower result than exact
calculation using total set of Feynman diagrams \cite{7}.
This fact also shows that measured number of events should be larger
than one predicted here.

The work was supported by the Russian Foundation for Basic Research
(Grant 98-15-96040) and Russian Ministry of Education (Grant
98-0-6.2-53).

\newpage

\begin{figure}[p]
\psfig{figure=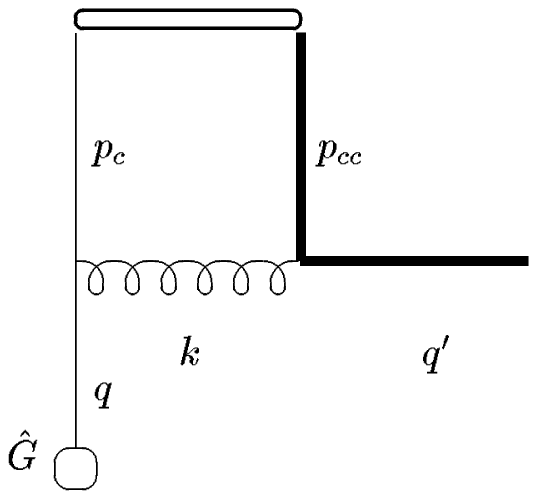,height=10cm,width=10cm,%
        bbllx=5cm,bblly=15cm,bburx=15cm,bbury=25cm}%
\vspace*{1.0cm}
\caption
{Diagram used for description of the c-quark to split into $\Oc$.}
\end{figure}

\begin{figure}[p]
\psfig{figure=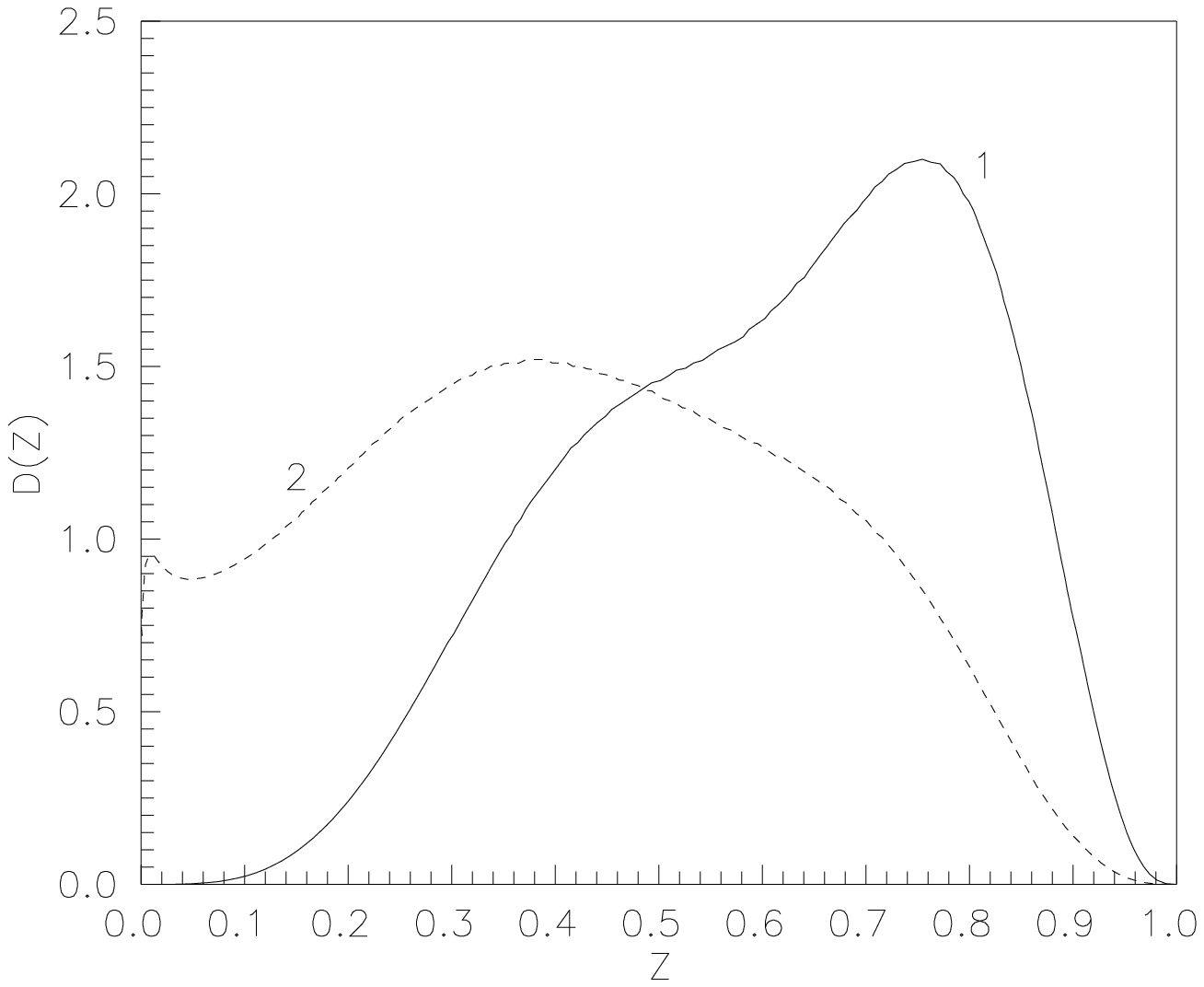,height=10cm,width=10cm,%
        bbllx=2cm,bblly=10cm,bburx=12cm,bbury=20cm}%
\vspace*{1.0cm}
\caption{
The fragmentation function $D_{c\to\Oc}(z,\mu)$ normalized to
unity at $\mu=\mu_0$ (curve 1) and $\mu=45$ GeV (curve 2).
}
\end{figure}

\end{document}